\documentstyle[pra,aps,multicol,graphicx,subfigure]{revtex}

\begin{document}


\title{The London Study of Vortex States in a Superconducting Film Due to a Magnetic Dot}

\author{Serkan Erdin\\  
Department of Physics, Northern Illinois University, DeKalb, IL, 60115\\
\& Advanced Photon Source, Argonne National Laboratory,\\
 9700 South Cass Avenue,  Argonne, IL, 60439}

\date{\today}
\maketitle

\begin{abstract}
Here we report a study of vortex states in a thin superconducting film 
with a 
magnetic dot grown upon it by means of a method based on London-Maxwell 
equations. Vortices with single quantum flux 
($\Phi_0 = h c 
/ 2 e$), giant vortices (  vortices with multiple
flux ) as well as antivortices ( vortices with 
negative vorticity ) are taken into 
consideration. It turns out that giant vortices occur, when 
the dot's size 
is sufficiently smaller ( $R \leq 4.5 \xi $) than the effective 
penetration depth $\Lambda$. 
In the case of a dot with sufficiently large size ( $R \geq 6 \xi $ ),    
the vortices with single quantum flux dominate
the vortex states. Their
 geometrical
patterns are predicted  up
to seven vortices. Our calculations do not show the spontaneous 
appearance of antivortices.
\end{abstract}  

\vspace{1cm}
                                                                                
\noindent PACS Number(s): 74.25.Dw, 74.25.Ha, 74.25.Qt, 74.78.-w
                                                                                
\vspace{1cm}

\begin{multicols}{2}

Type II superconductors are used in a wide variety of technological 
applications 
due to their high critical currents and fields \cite{tinkham}. In these 
superconductors,  vortices appear when the 
magnetic 
field exceeds the first critical field $H_{c1}$. Under  external current 
or 
field, 
vortices move, which causes the superconductor to switch to a resistive 
state. As a result, the system  loses its superconductivity. Because of 
this, vortex 
pinning is quite important in applications of type II superconductors. One 
of 
the ways to pin vortices is to use  the magnetic subsytems with either 
out-of-plane or in-plane magnetization. These subsystems are capable of  
trapping   both 
vortices and antivortices, depending on the orientation  and strength of 
their  
magnetization. The aforementioned systems are not only important  for  
technological applications, such as 
devices that 
can be tuned by weak magnetic fields, but also offer rich physical effects 
which are not observed in the individual parts. Some of these effects were 
predicted elsewhere \cite{pok1,domains,se1,se2,pok2,pok3}.

In the recent decade, magnetic 
dots growing on top of SC films have been extensively studied both 
experimentally 
\cite{exp} and 
theoretically \cite{pok-rev,serk-rev}. In experimental studies, magnetic 
dots with in-plane 
magnetization are fabricated from Co, Ni, Fe, Gd-Co and Sm-Co alloys, 
whereas, 
for 
the dots with magnetization perpendicular to the plane, Co/Pt multilayers 
are used \cite{vanbael}. These studies report commensurability effects on 
transport 
properties, which comfirm that the dots create and pin vortices. On 
theoretical side, several realizations of the aforementioned systems are 
analyzed  through the Landau-Gizburg framework and London theory. In these 
works, the authors investigated
the conditions for vortices to appear and calculated their 
geometric 
configurations in equilibrium. 

Recently, Priour {\it et al.} studied the  vortex states of a SC film with 
a  
magnetic dot array grown upon on it, through Ginzburg-Landau theory, and 
found several different configurations of vortices in the SC film \cite{priour}. 
Similar study was also done by Peeters {\it et al.}, but in the presence 
of a 
single 
dot \cite{marm,peeters2}. One of the interesting results that is 
reported 
by these 
works is, when the dot's size is on the order of a few coherence lengths 
$\xi$, vortices  with multiple flux quanta ( the giant vortices )  
appear. The giant vortices  are previously studied in the 
Landau-Ginzburg 
framework \cite{giant}. The stable antivortex (AV) 
( vortices with vorticity opposite to vorticity of those
trapped under the dot )
states were reported in similar systems \cite{peeters3}. The most 
interestingly, Kanda {\it et. al.} recently reported the experimental evidence 
of the GV states and the other rearrangements of vortices in a mesoscopic 
SC disc, using the multiple-tunnel-junction method \cite{kanda}.

Earlier, we studied the vortex states due to 
a ferromagnetic (FM) dot with out-of-plane ( perpendicular to the film ) 
magnetization on top 
of a
SC thin film \cite{common}. In that work, we limited ourselves only to the 
vortices trapped 
under the dot, and calculated the equilibrium configurations, up to three 
vortices.  However, in a more realistic picture, both vortices under the 
dot and 
antivortices  outside the dot's boundaries can
appear spontaneously, because 
total flux due to the magnetic dot over the entire infinite SC film is 
zero. Actually, this problem was 
studied for magnetic dipoles with out-of-plane magnetization 
\cite{milosevic}.
Another interesting case is giant vortex (GV) states  with multiple unit 
flux 
$\Phi_0$. In usual circumstances, it is
hard to get vortices with multiple flux quanta, because their energy grows
as the square of their vorticity. However,
this energy cost can be overcome by the dots with radii of  sufficiently 
less
than effective penetration depth $\Lambda=\lambda^2/d_{sc}$
\cite{abrikosov}, where $\lambda$ is the London penetration
depth and $d_{sc}$ is the thickness of the SC film,  and with
sufficiently large magnetization.

In this article, we focus on the complete picture together with antivortices and giant 
vortices, and aim to get analytical insight for the spontaneous vortex, 
antivortex and giant vortex states. 
For this purpose, we
pursue the method based on London-Maxwell equations, which
is fully developed elsewhere \cite{common}. Though the London
approach
works
well   at the  high $\kappa = \Lambda/\xi$ limit, 
it perfectly 
serves our purpose
and enables us to do  analytical calculations. 
 In our analytical and 
numerical calculations, we show that there are three regimes for the 
vortex states according to  the dot's size. Namely, when the 
dot's size is sufficently small, the GV states are dominant. 
If the dot's 
size is sufficently large, we see only vortex states with single quantum flux (SQF). 
There is also 
an intermediate region in which both GV and SQF 
states 
appear. 
We  calculate the geometrical configurations of SQF states in 
equilibrium up to seven vortices. In addition, our calculations do not 
show 
any AV states.  At this
point, we need to remark that  the London framework might seem
unreliable to some readers,
when the ratio
of the dot's size to the coherence length, $\xi$,   is in the order of
unity.
However, one might expect that the vortices in the London framework,
usually prefer staying in SQF states rather than GV
states, since the London theory treats the vortex core as a point. Our
approach then underestimates the dot's size, when the GV states appear.
When the vortex cores are treated more accurately, as in Landau-Ginzburg
framework, one might expect that the dot's size would be rather larger
than that we estimate in this work. Nevertheless, the physics and
qualitative result  is still captured correctly in the London theory,
although the
results might be quantitatively different.

In the following section, we briefly 
introduce the method we follow in this article. Next, we study the giant 
vortex states with and without antivortices. The third section is devoted 
to 
a discussion on the SQF vortex states with and without AV states. The last 
section consists of the discussions and  conclusions. 


\section{Method}

\begin{figure}[htb]
\begin{center}
\includegraphics[angle=0,totalheight=2.0in]{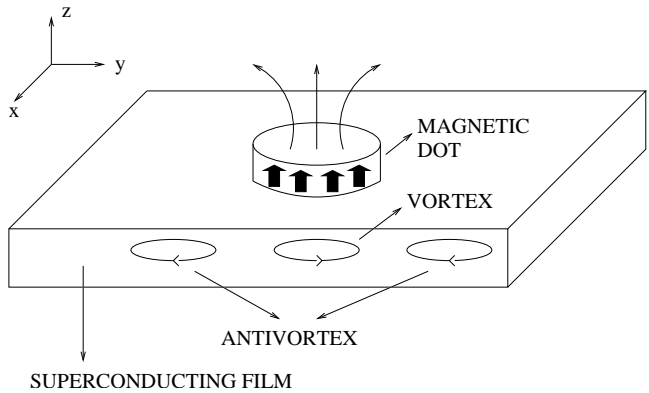}
\caption{\label{fig:mdot}Magnetic dot on a superconducting film. }
\end{center}
\end{figure}

In order to attack  this problem, we consider a thin circular magnetic 
dot of radius $R$ with magnetization perpendicular to its plane at $z=d$, 
placed upon a 
thin SC 
film which both lay on the x-y plane $z=0$ (see Fig. \ref{fig:mdot}). The 
dot's 
magnetization reads
\begin{equation}
{\bf m} = m \Theta( R - r) \delta ( z - d) \hat z. 
\label{magdot}
\end{equation}

The energy of the system in the presence of many vortices 
with
arbitrary
vorticities $n_i$ is described 
elsewhere \cite{common}. The appearance of N
vortices with arbitrary positions ${\bf r}_i$ in the system changes
the energy by an amount:
\begin{eqnarray}\label{change}
  \Delta_N &=& \sum_{i=1}^{N} n_i^2 \varepsilon_v +
\frac{1}{2} \sum_{i\neq j}^N n_i n_j \varepsilon_{vv} (r_{ij})
\nonumber \\ 
&+& 
\sum_{i=1}^N
n_i \varepsilon_{mv}( r_i).
\end{eqnarray}
Here $\varepsilon_v = \varepsilon_0\ln (\Lambda /\xi)$ is the
self energy of a vortex without a magnetic dot, $\varepsilon_0 =
{\Phi^2_0 /(16 \pi^2 \Lambda)}$, $\varepsilon_{vv}$ is the
vortex-vortex interaction, and $\varepsilon_{mv}$ is the vortex-magnetic
dot interaction. The vortex-vortex interaction reads 
\begin{equation}
\varepsilon_{vv} (r_{ij}) = \frac{\varepsilon_0}{\pi} \left [ H_0
\left (\frac{r_{ij}}{2
\Lambda} \right ) - Y_0 \left (\frac{r_{ij}}{2  \Lambda} \right )\right 
],\label{evv}
\end{equation}
\noindent where $r_{ij} = |{\bf r}_j - {\bf r}_i|$ , $H_0 ( x )$ and $Y_0
( x )$ are the Struve function of the zeroth order and the modified Bessel
function of the second kind of the
zeroth order, respectively \cite{abrom}.  The asymptotics of
Eq.(\ref{evv}) are
\begin{eqnarray}
\varepsilon_{vv} &\approx& 2 \varepsilon_0 \sum_{i > j} n_i n_j \ln
\frac{\Lambda}{|{\bf r}_i - {\bf r}_j|},
|{\bf r}_i - {\bf r}_j| < \Lambda \nonumber \\
 &\approx&  4 \varepsilon_0 \Lambda \sum_{i > j} n_i n_j \frac{1}{ |{\bf 
r}_i -
{\bf r}_j| }, |{\bf r}_i - {\bf r}_j| >  \Lambda.
\label{evv-asym}
\end{eqnarray}
Note that in our further calculations, core energy $\varepsilon_{core}
\sim
0.809 \varepsilon_0$ is renormalized to $\Lambda$ in  logarithmic terms.
The vortex-magnetization interaction energy $\varepsilon_{mv}$ is 
represented by 
\begin{equation}\label{epsilon-mv}
  \varepsilon_{mv} (r_i) =
  -m\Phi_0 R\int_0^{\infty}\frac{J_1(qR) J_0(qr_i) e^{-q d}dq}{1+2\Lambda 
q}.
\end{equation}
When $d\neq0$  and is significantly smaller than $R$, the correction to 
the
$\varepsilon_{vm}$ is on the order of
$d/R$. Throughout the paper, we stick to  $d<<R$ approximation to limit 
the
number of parameters in our
analytical calculations, which does not change
the qualitative results. In our numerical calculations, however, we 
take $d\neq
0$ to make sure that numerical integrations converge.
The asymptotics at $d =0$  are
\begin{eqnarray}
\varepsilon_{vm} &\approx& - m \Phi_0 R \sum_i n_i \left ( \frac{1}{2 
\Lambda} -
\frac{r_i^2}{8 \Lambda R^2} \right ), r_i < R < \Lambda \nonumber \\
 &\approx& - m \Phi_0 R \sum_i n_i \left ( \frac{1}{R} -
\frac{3 r_i^2 \Lambda }{2 R^4} \right ), r_i  < \Lambda < R \nonumber \\
&\approx& - m \Phi_0 R \sum_i n_i \frac{R^2 \Lambda}{r_i^3}, R < \Lambda <
r_i \nonumber \\
&\approx& - m \Phi_0 R \sum_i n_i \frac{R^2}{4 \Lambda r_i}, R < r_i <
\Lambda. \label{evm-asym}
\end{eqnarray}
 Additionally, 
under the assumption that the magnetization does not change direction due
to its interaction with vortices, the self-magnetization energy is 
irrelevant to our
calculations. Because of this, we do not take this energy into account.  
In the rest of the paper, $\Delta_N$ denotes the energy of  $N$ vortices 
with single quantum flux ($L=N$ state), while $\Delta_N^g$ represents the 
energy of a giant vortex with vorticity $N$ ( $L_g=N$ state ). 
$\Delta_N^{ga}$ is used for a GV with $N$ vorticity and $N$ SQF antivortex 
around it, whereas $\Delta_N^a$  is used for $N$ SQF states along with $N$ 
AV states  ($L_a= N$ state).

\section{The Giant Vortices}
The self energy of a GV with vorticity $n$ is proportional to its 
vorticity's square
$n^2$, whereas the self energies of $n$ vortices with SQF is 
just proportional to $n$. Because of this, 
GVs are unstable in usual superconductors. Even if they pop up,
 they decay into vortex states with vorticity $n=1$.
To get an idea about when GV states might occur,  let us consider the 
following scenario in which   
SQF vortex states are trapped in a small region  through a 
sufficiently small-sized magnetic dot so that they cannot go to 
further distances.
In order for SQF states to be stable, the system requires more gain in 
energy, since   energy grows due to the repulsion between SQF states
when they get closer.  For this purpose, the dot is required to have 
sufficiently high magnetization. 
We  then expect the GVs can be stable if the dot's size is small enough and its magnetization is high enough. These physical
arguments are our starting point. In this section,
 we consider a small circular dot $R<\Lambda$, placed upon a SC thin film.
Previously, we studied the vortex states in the absence of 
antivortices and found that at first the  $L=1$ state appears, and, with 
further increase of  magnetization, $L=2$,$L=3$ states occur in 
turn. In addition, the number of states depends on the magnetization and 
the dot's size. We 
anticipate  the same sitution when the dot's 
size is small  enough to make the GV states stable. Contrary to our 
previous results, we expect $L=1$, $L_g=2$, $L_g=3$ and so on, when only 
GV 
states are present. 
The first question we ask here is how small does the dot have to be so 
that the $L_g=n$ state will be preferred over the  $L=n$ state. The 
necessary 
conditions for this case are                                                       
$\Delta_n^g < \Delta_n$ and $\Delta_n^g < 0$. 
We first evaluate this problem analytically, and, in doing so,  we 
disregard the antivortices to simplify the calculations. They will be 
discussed at the end of this section.  
At the  $R<\Lambda$ limit, the effective energy of a GV state $L_g=n$ 
reads
\begin{equation}
\Delta_n^g \approx n^2 \varepsilon_0 \ln\frac{\Lambda}{\xi} - \frac{n m
\Phi_0 R }{ 2
\Lambda},\label{engiant}
\end{equation}
whereas the energy of $L=n$ state is 
\begin{eqnarray}
\Delta_n &\approx& n  \varepsilon_0 \ln\frac{\Lambda}{\xi} + 2 
\varepsilon_0
\sum_{i>j}
\ln \left ( \frac{\Lambda}{|{\bf r}_i-{\bf r}_j|} \right )  
\nonumber \\
&-& \sum_{i}^n \frac{ m
\Phi_0 R
}{
2 \Lambda}  \left ( 1 - \frac{r_i^2}{4 R^2} \right ), \label{nvor}
\end{eqnarray}
where $r_i$ is the position of $i$th antivortex from the dot's center.
This problem can be simplified by using circular symmetry in the system.
If the vortices are considered to be situated on a ring of
radius $r_v>\Lambda$ such that nearest vortices are equally distant from
each
other, the distance between $i$th and $j$th antivortices can be expressed
as  $|{\bf r}_i - {\bf r}_j| = 2 r_v | \sin ( \pi (i-j)/n) |$. Thereafter,
taking the radius of the ring as
$r_v$, the sum in the second term of Eq.(\ref{nvor}) can be replaced by
$\sum_{k=1}^{n-1} ( n - k)
\ln (\Lambda/(2 r_v \sin(\pi k /n)))$. Under this assumption,
the last term of  Eq.(\ref{nvor}) becomes  $(n  m
\Phi_0 R /
2 \Lambda) ( 1 - r_v^2 /(4 R^2) )$. Note that this assumption is 
valid for up to a few vortices. We expect that new rings occur as new 
vortices come out. In the next section, we show that the first ring 
carries 6 vortices on it.  
However this  suffices for our analysis, since 
it  is very unlikely to see a GV state with vorticity $n=6$.  
As 
a result, the effective energy is
now a function of $r_v$. Optimizing the effective energy with respect to  
$r_v$, we
find its equilibrium value  as
\begin{equation}
r_v = \sqrt{4 ( n - 1 ) \frac{\varepsilon_0 R \Lambda}{m 
\Phi_0}}.\label{rv}
\end{equation}
Directly substituting Eq.(\ref{rv}) into Eq.(\ref{nvor}), we find
\begin{eqnarray}
\label{nvorop}
\Delta_n &=&  n \varepsilon_0 \ln\frac{\Lambda}{ \xi} + \frac{n ( n - 1
)}{2} \varepsilon_0 \ln \left ( \frac{\Lambda m \Phi_0  e}{1 6 ( n - 1 ) 
R\varepsilon_0} \right ) \\ \nonumber 
&-& 2 \varepsilon_0 \ln C_n - \frac{n m \Phi_0 R }{
2 \Lambda},
\end{eqnarray}
where $C_n = \Pi_{k=1}^{n-1} \sin ( \pi k / n )^{n - k}$ and $e=2.71828$. Let us compare
Eq.(\ref{nvorop}) with the energy of a system in the presence of a GV  
with vorticity $n$
\begin{equation}
\Delta_n^g \approx n^2 \varepsilon_0 \ln\frac{\Lambda}{\xi} - \frac{n m
\Phi_0 R }{
2 \Lambda}.\label{engiant}
\end{equation}
From Eq.(\ref{engiant}) and Eq.(\ref{nvorop}),
we find
\begin{equation}
\frac{R}{\xi} < \frac{1}{\kappa} \frac{m \Phi_0 }{\varepsilon_0} \frac{e}{16 ( n
- 1 ) C_n^{\frac{4}{n ( n - 1 )}}}. \label{rcrit1}
\end{equation}
The above equation gives the upper limit for the dot's size as a function of magnetization of vorticity.
When the dot's size is  smaller than the upper limit, at first the $L=1$ 
state appears. The $L_g =2$ state follows next. 
With a further increase of magnetization, new GV states appear 
spontaneously. This picture is very similar to our previous
results, except we have the GV states instead of the SQF states. In order 
for 
the $L_g=n$ state to appear, the necessary conditions are 
$\Delta_n^g<\Delta_{n-1}^g$ and $\Delta_n^g<0$. These conditions give us 
the lower limit as
\begin{equation}
\frac{R}{\xi} > \frac{2( 2 n - 1)\kappa \ln \kappa}{\frac{m 
\Phi_0}{\varepsilon_0}}.\label{rcrit2}
\end{equation}

\end{multicols}

\begin{figure}
\centering
\subfigure[$L_g=2$  vortex state] 
{
    \label{fig:gvs:a}
    \includegraphics[angle=270,width=8cm]{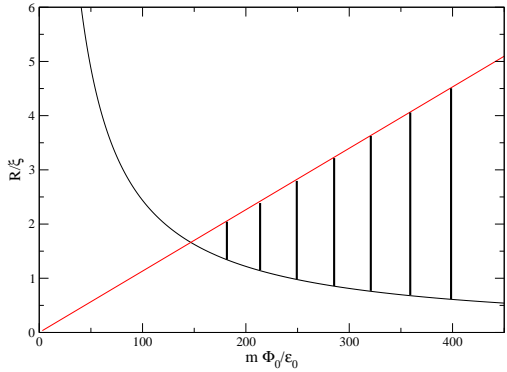}
}
\hspace{1cm}
\subfigure[$L_g=3$ vortex state] 
{
    \label{fig:gvs:b}
    \includegraphics[angle=270,width=8cm]{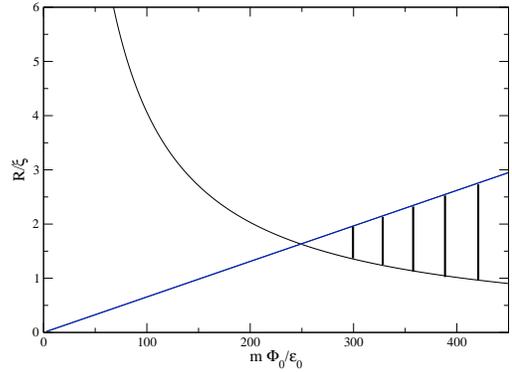}
}
\caption{The shaded regions show where $L_g=2$ and $L_g=3$ states can 
appear.}
\label{fig:gvs} 
\end{figure}

\begin{multicols}{2}

\begin{figure}
\centering
\subfigure[$R/\Lambda=0.3$] 
{
    \label{fig:gven:a}
    \includegraphics[angle=0,width=6cm]{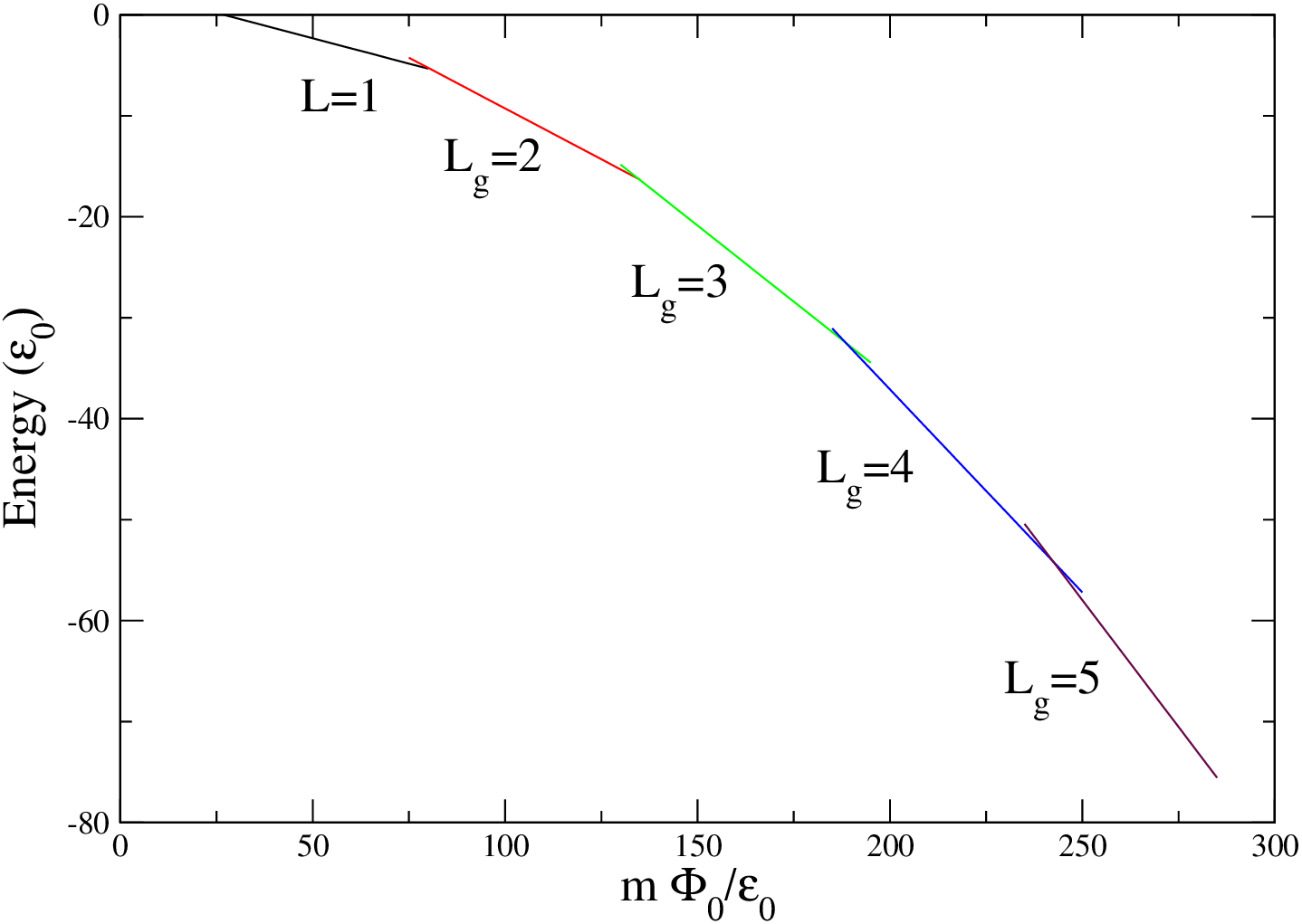}
}
\hspace{1cm}
\subfigure[$R/\Lambda=0.35$] 
{
    \label{fig:gven:b}
    \includegraphics[angle=0,width=6cm]{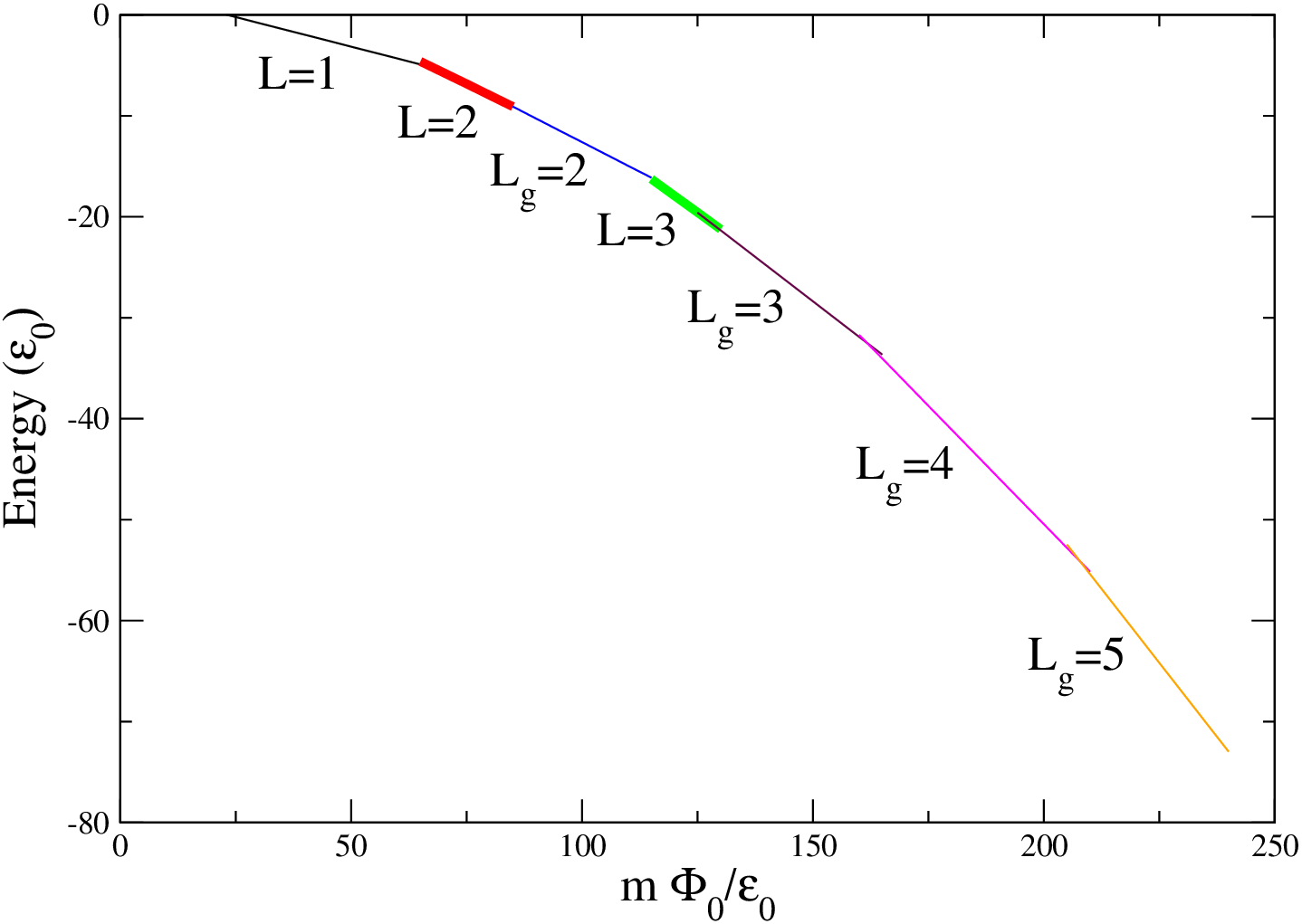}
}
\vspace{1cm}
\subfigure[$R/\Lambda=0.4$] 
{
    \label{fig:gven:c}
    \includegraphics[angle=0,width=6cm]{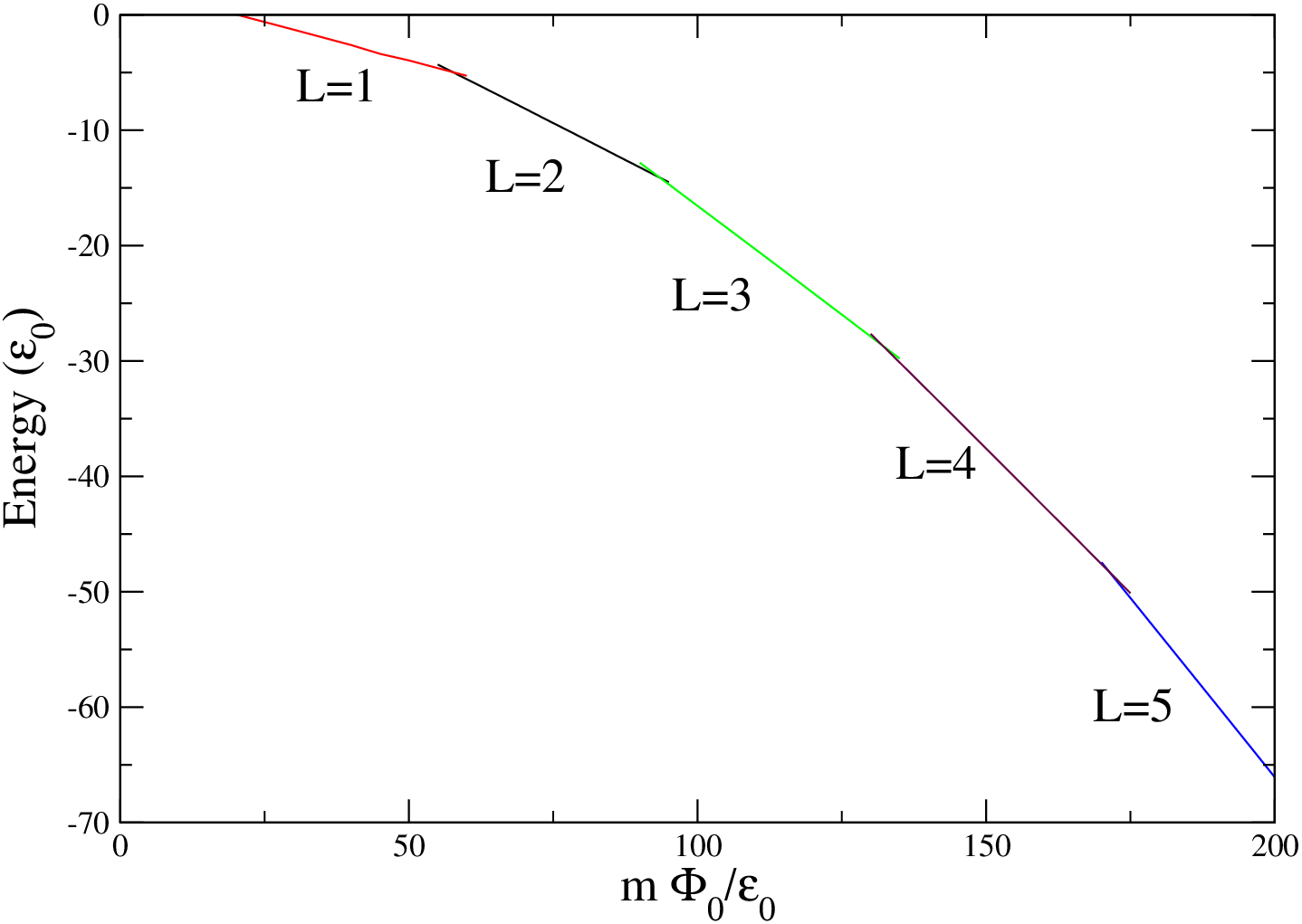}
}
\caption{\label{fig:gven}The vortex states when $R/\Lambda = 
0.3,0.35,0.4$.}
\end{figure}

\begin{figure}
\begin{center}
\includegraphics[angle=0,totalheight=3.0in]{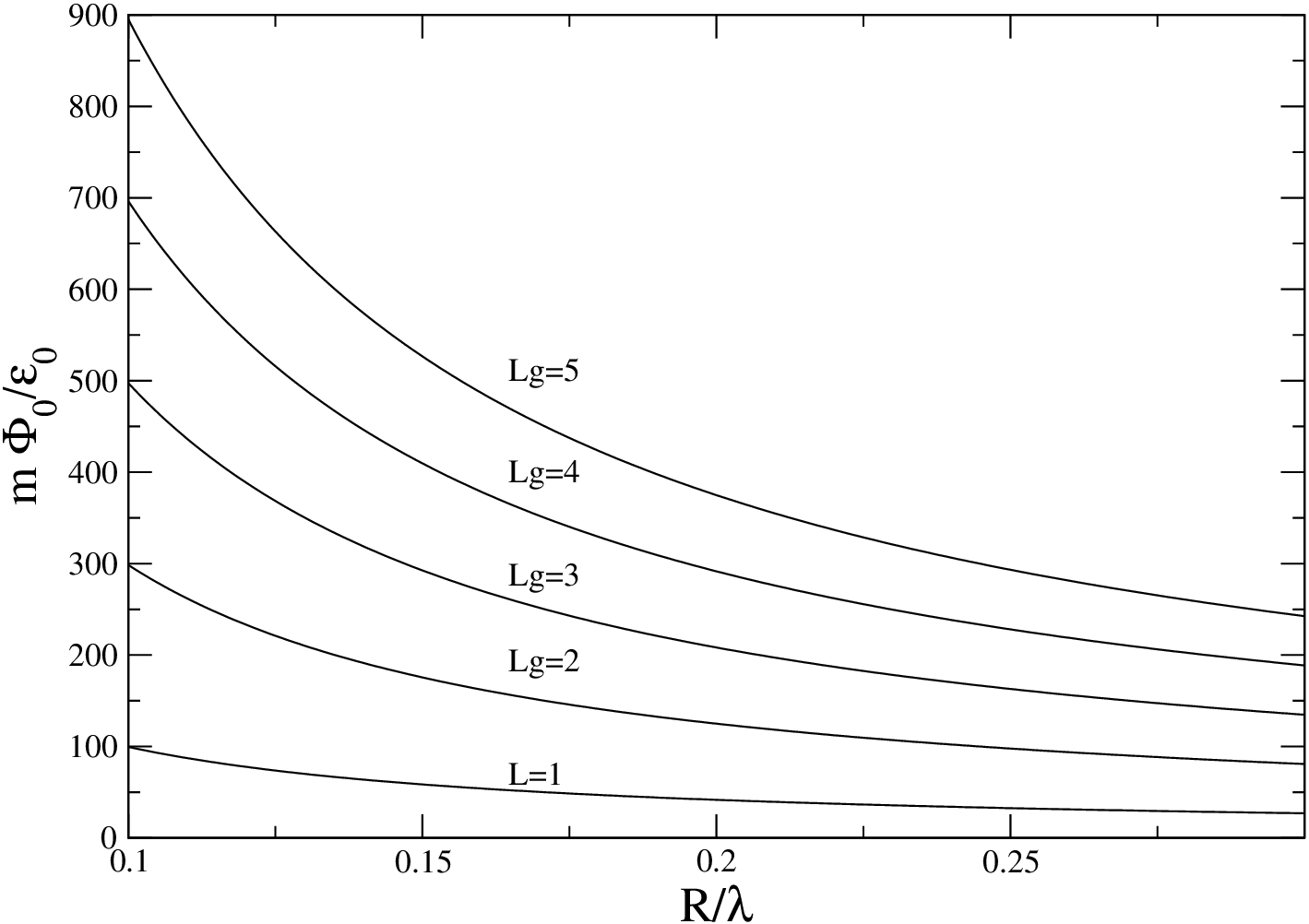}
\caption{\label{fig:gvphase}Phase diagram of GV states. }
\end{center}
\end{figure}

As seen from Fig. \ref{fig:gvs}, as the vorticity  of the GV states 
increases, the region in which they might appear shrinks. This 
suggests that the high vorticity GV states are more unlikely in usual 
circumstances.   As a result of analytical
calculations, we expect that the giant vortices can  occur when $R/\xi \sim 3,4$. To get the better 
result, we do numerical calculations for small values of $R/\Lambda$ as 
analytical calculations suggest. In our numerical calculations, 
we set  $\kappa = 15$ and $d/\Lambda = 0.05$. The results of the numerical 
calculations are depicted in Fig. \ref{fig:gven}. We find that 
the giant vortices dominate the vortex states when $R/\Lambda = 0.3$ 
($R/\xi = \kappa R/\Lambda = 4.5$).
However, along with the further increase of the dot's size to $0.35$ ( 
$R/\xi = 5.25$), we 
begin to see $L=2$ and $L=3$ states. However, we do
not see $L=4$ or other states with a single quantum flux. The physical 
reason 
for 
this is that the dot is still small enough so that the vortices can find 
the enough space to decay into SQF states. When the size is increased to 
0.4 ($R/\xi = 6$)
,
 we do not see any GV states anymore. The phase diagram of GV states is 
given  in Fig. \ref{fig:gvphase}. The very bottom curve separates the  no 
vortex 
region ( area below the curve ) from the region where the $L=1$ state 
becomes 
stable ( area between the 1st and 2nd curves).  This curve is 
calculated from the equation $\Delta_1=0$. Next,  the $L_g=2$ state 
follows. 
The 
corresponding curve, which separates the $L=1$ and $L_g=2$ states ( 
area 
between 2nd and 3rd curves ), is found 
from $\Delta_2^g=\Delta_1$. The following transitions are found from the 
equation
$\Delta_n^g = 
\Delta_{n-1}^g$ for $n=2,3  ...$.

\end{multicols}

\begin{multicols}{2}


Next, we consider the possibility of AV states together with GV states. 
The total flux due to the magnetic dot is zero throughout the entire 
SC film. The flux due to the dot causes the spontaneous appearance of GV 
or 
SQF states with the dot boundaries. By the same token, the flux due to 
the dot outside the dot boundaries might cause spontaneous 
AV states, when the dot's magnetic field penetrates into 
the the film.   
Since the total flux 
is zero over the entire SC film, we expect then the number of 
antivortices with single quantum flux equals  the GV states's vorticity. 
In principle, the total GV's vorticity  might differ from the number of 
SQF AV 
states in such a way that the net vorticity is positive. For the sake of 
simplicity, we do not consider this case. However, if stable AV vortices
do not exist, this scenerio is quite unlikely.  Because of 
circular symmetry, we assume that they are located  on a ring of 
radius $R<r_a<\Lambda$. Again, we stress that this assumption is valid 
only for a few 
vortices. It is, however, sufficient to get a  
qualitative idea 
about AV states. Under the assumption 
that AV  states are located within 
the region of $\Lambda$ ($r_a < \Lambda$ limit),   the effective energy 
of a GV state $L_g=n$ with n antivortices 
surrounding it reads 

\begin{eqnarray}
\label{antvor}
\Delta_n^{ga} &\approx& ( n^2+n) \varepsilon_v + n(n-1) \varepsilon_0 \ln 
\frac{\Lambda}{2 r_a} -
2 n^2 \varepsilon_0 \ln \frac{\Lambda}{r_a} \nonumber \\ 
&-& 2 \varepsilon_0 \ln C_n - \frac{n m \Phi_0 R}{2 \Lambda} \left ( 1 - 
\frac{R}{2 r_a} \right ). 
\end{eqnarray}
Minimizing the above equation with respect to  the ring's radius $r_a$, we 
find 

\begin{equation}
r_a = \frac{m \Phi_0 R^2}{4 ( n +1 ) \varepsilon_0 \Lambda}.\label{ra}
\end{equation} 
Directly substituting  Eq.(\ref{ra}) into Eq.(\ref{antvor}), we obtain

\begin{eqnarray}
\Delta_n^{ga} &=& n(n+1)\varepsilon_0\ln \left(\frac{2 \kappa e  
R^2 
m\Phi_0}{(n+1) 
\Lambda^2 \varepsilon_0}\right ) - 2 \varepsilon_0 n^2 \ln 2 \nonumber \\
&-& 2 \varepsilon_0 \ln C_n - 
\frac{n m 
\Phi_0 R}{2 \Lambda}.
\end{eqnarray}
The GV state $L_g=n$  is more energetically favorable than the $L_{ga} =n$ 
state, when $\Delta_n^{ag} < \Delta_n^{g}$ and $\Delta_n^{ag} < 0$. Using the  
initial approximation $R<r_a<\Lambda$ and Eq.(\ref{ra}), we determine the boundaries of 
$m \Phi_0/\varepsilon_0$ value as
\begin{equation}
4 ( n + 1 ) \frac{\Lambda}{R} <         
\frac{m \Phi_0}{ \varepsilon_0} < 4 ( n + 1 ) \frac{\Lambda^2}{R^2}.\label{range}  
\end{equation}
Next, we compare $\Delta_{n}^{ga}$ and 
$\Delta_{n}^{g}$ for several values of vorticity $n$ within the limits of 
$m \Phi_0/\varepsilon_0$ which are given in Eq.(\ref{range}).
It turns out than the giant vortices are always more 
favorable 
than the case in which they appear together with antivortices in 
$R < r_a<\Lambda$ limit. At the  opposite limit, $r_a > \Lambda$ , it is 
quite 
unlikely to have stable states together with antivortices. Since 
the effective energy goes to $n ( n + 1 ) \varepsilon_0 \ln \kappa - n  
m \Phi_0 R/(2 \Lambda)$ when $r_a >> \Lambda$. The energy difference $n 
\varepsilon_0 \ln \kappa$ can be overcome only when $R \sim 
\xi$. However, we do not consider that case in this work  because of the 
inaccuracy of London 
approximation at the  small $\kappa$ limit. Actually, when 
Eq.(\ref{change})  is 
optimized 
together with antivortices, it manifests instability for vortex states, 
that is, high vorticity states become more favorable, and no low 
vorticity states 
appear. The reason for this is logarithmic divergence in the $H_0 ( x 
) - Y_0 ( x)$ function. As $x \rightarrow 0$, this function goes 
logarithmically to infinity, which suggests that in the London 
approximation, a vortex-antivortex interaction is not taken care of 
accurately when the  distance between the vortex-antivortex pairs is on 
the 
order 
of a few $\xi$. For this situation, 
we believe that the best approach is the nonlinear Landau Ginzburg 
equation.

%

\section{SQF Vortex States}

In the second section, we found that, when the dot's size is as large as 
about $0.4 \Lambda$ ($6 \xi$), only  SQF vortex states occur. We devote 
this section 
to the analysis of geometrical patterns formed by SQF states and their 
phase diagram. 
In this part, we 
focus on two possible cases; vortices with and without antivortices. The 
former case  has been  previously analyzed for up to three 
vortices \cite{common}. We determined the  geometrical configurations of 
vortices in the ground state.  Due to
symmetry, the centers of the two vortices are located on a
straight line connecting the vortices with the center of the dot
at equal distances from the center.
The occurrence of two vortices can be experimentally
detected as a violation of circular symmetry of the field. For
 three
vortices, the equilibrium configuration is a regular
triangle. In this section, we consider further cases. Due to the circular 
symmetry, vortices seem to favor location on a circle such that they are 
equally distant from nearest neighbors. However, there is another possible 
case in which a vortex is situated at the dot's center while others are 
located on a circle as  described just above, in the absence of 
antivortices.

\begin{figure}[htb]
\begin{center}
\includegraphics[angle=0,totalheight=2.5in]{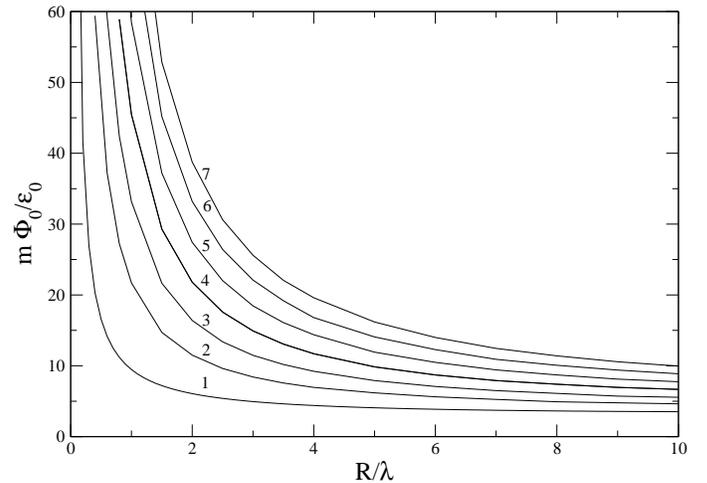}
\caption{\label{fig:vphase}Phase diagram of the  vortex states. }
\end{center}
\end{figure}

\begin{figure}
\centering
\subfigure[One  vortex] 
{
    \label{fig:vstates:a}
    \includegraphics[width=3cm]{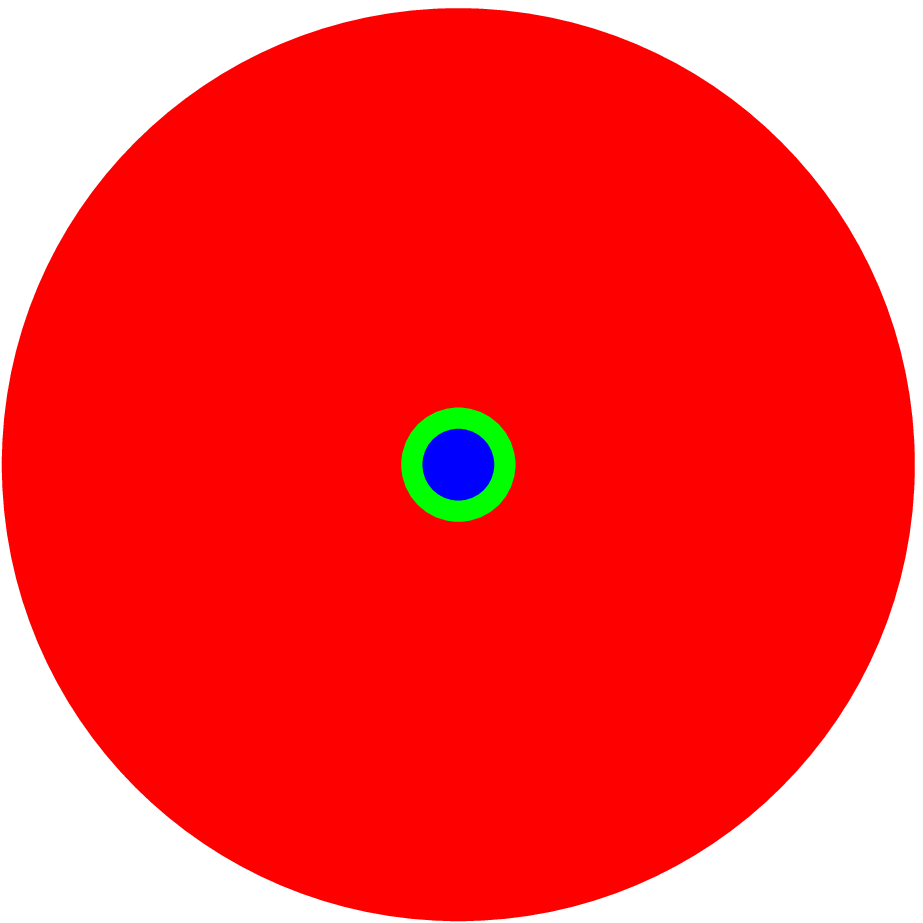}
}
\hspace{1cm}
\subfigure[Two vortices] 
{
    \label{fig:vstates:b}
    \includegraphics[width=3cm]{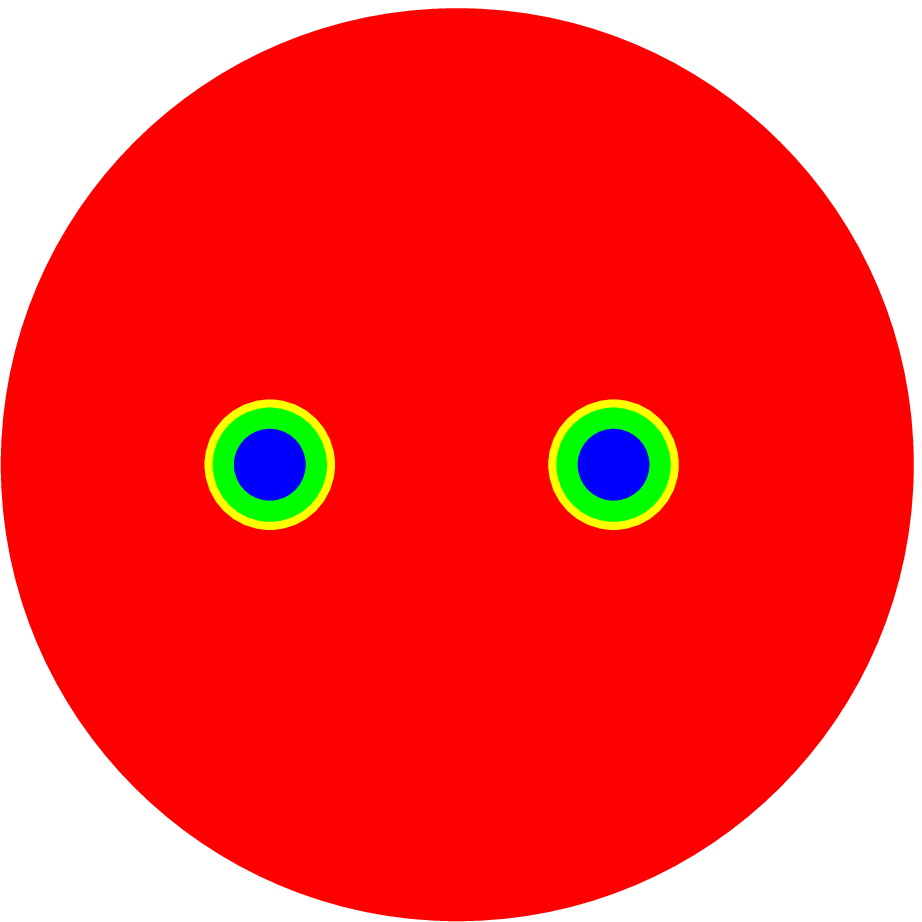}
}
\hspace{1cm}
\subfigure[Three vortices] 
{
    \label{fig:vstates:c}
    \includegraphics[width=3cm]{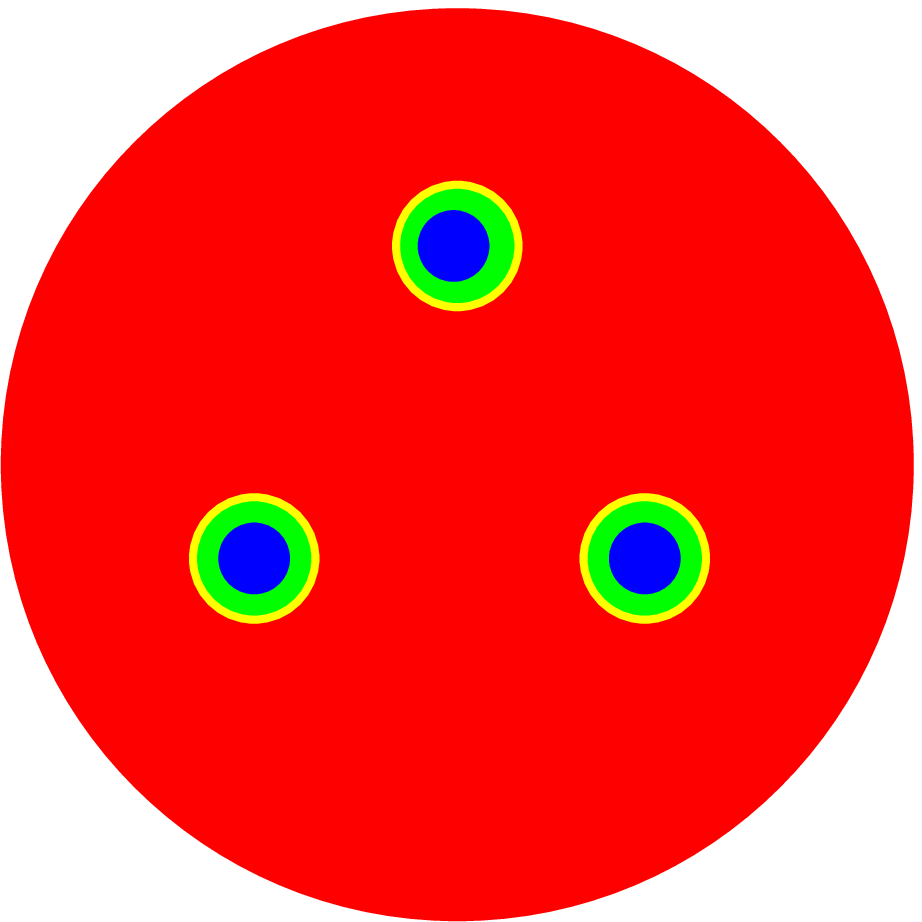}
}
\hspace{1cm}
\subfigure[Four vortices] 
{
    \label{fig:vstates:d}
    \includegraphics[width=3cm]{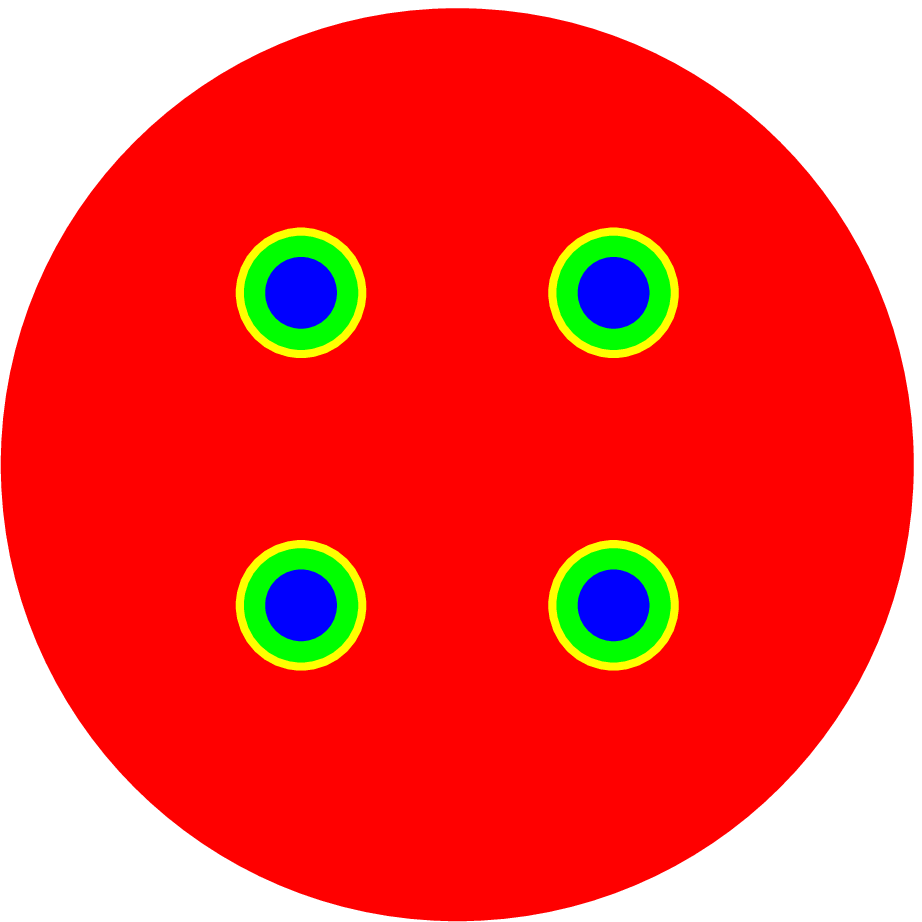}
}
\vspace{1cm}
\subfigure[Five vortices] 
{
    \label{fig:vstates:e}
    \includegraphics[width=3cm]{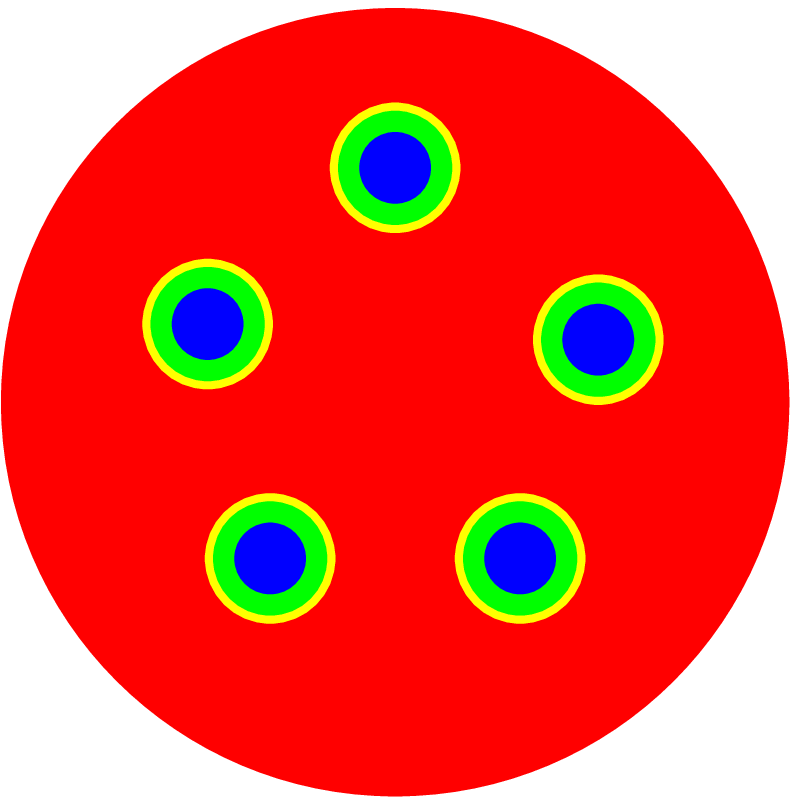}
}
\hspace{0.5cm}
\subfigure[Six vortices] 
{
    \label{fig:vstates:f}
    \includegraphics[width=3cm]{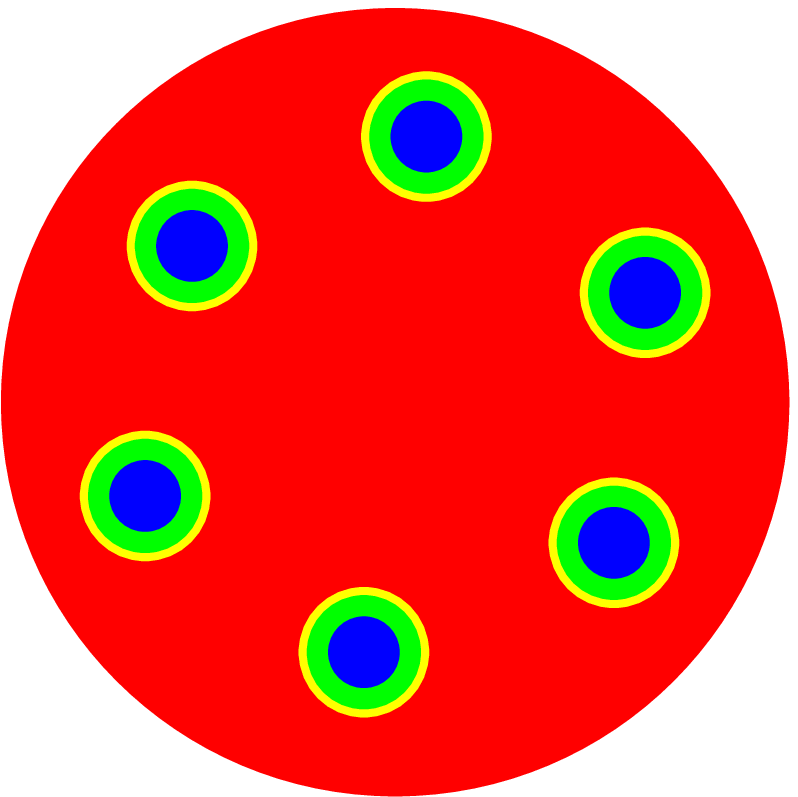}
}
\vspace{1cm}
\subfigure[Seven vortices] 
{
    \label{fig:vstates:e}
    \includegraphics[width=3cm]{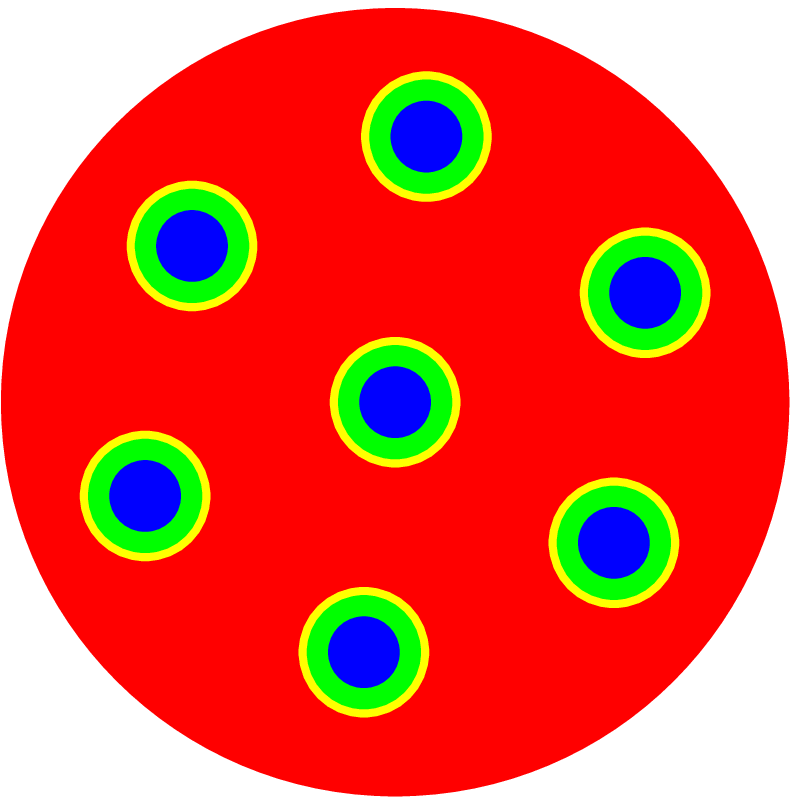}
}
\caption{Vortex states}
\label{fig:vstates} 
\end{figure}

In order for N vortices to appear, the necessary condition is that 
$\Delta_N < 0$ and $\Delta_N < \Delta_{N-1}$. Using this criteria, we can 
determine in what configurations and order the vortices appear. To this 
end, we study only vortices with positive vorticity that are situated 
under the dot. The next step is to minimize with respect to the positions 
of vortices. We first start with one vortex. It turns out that it appears 
at the center of the dot. $\Delta_1$ is a function of two dimensionless 
parameters $m \Phi_0/\varepsilon_0$ and $R/\Lambda$. $\Delta_1$ defines a 
critical curve that separates regions with or without vortices, as is 
depicted in Fig. \ref{fig:vphase}.  Next, we calculate $\Delta_2$  for two 
vortices. Our 
calculations show that they are located on a straight line 
connecting the vortices with the center of the dot at equal distances from 
the center. The equation $\Delta_2=\Delta_1$ gives the second curve in 
Fig. \ref{fig:vphase}, 
which separates the regions where one vortex and two vortices appear. That 
is, the area between the first curve and second curve indicates where one 
vortex 
occurs, while the region between the second curve and third curve is for 
two 
vortices. The equilibrium configuration of three vortices is a regular 
triangle. A
further increase of   $m \Phi_0/\varepsilon_0$
makes other  vortex states more energetically favorable. Up to seven 
vortices  tend to locate on a ring and equally distant from each 
other. Namely, four vortices form a square, 
whereas five 
vortices appear 
at 
the corners of a pentagon. The equilibrium configuration for six vortices 
is a hexagon. However, the seventh vortex appears at the dot's center, 
while the  other 
six vortices form a hexagon around it. Geometrical patterns of up to seven 
vortices are depicted in Fig. \ref{fig:vstates}. In principle, there 
exists an 
infinite series of such transitions. Here, we limit ourselves to seven 
vortices. The curves in a phase diagram of these seven vortices are 
obtained  from 
$\Delta_N=\Delta_{N-1}$ equation, which is a function of $R/\Lambda$ 
and $m \Phi_0/\varepsilon_0$. Fixing the former, we calculate the latter 
variable. 
We do this for various values of  $R/\Lambda$. 
Fitting the points that are obtained from  numerical calculations, we 
find 
the generic function 
$m \Phi_0/\varepsilon_0 \approx 4.22 N/(R/\Lambda)^{\nu}$ where $\nu = 
0.99 \pm 0.03$. 

\vspace{15pt}

\end{multicols}

\begin{multicols}{2}

Equilibrium positions of SQF states  change significantly with increasing 
$m \Phi_0/\varepsilon_0$. That is, the larger $m \Phi_0/\varepsilon_0$, 
the more  vortices are pushed towards the dot's center.  

\begin{figure}
\centering
\subfigure[$L_a=2$  vortex state] 
{
    \label{fig:anti:a}
    \includegraphics[width=3.5cm]{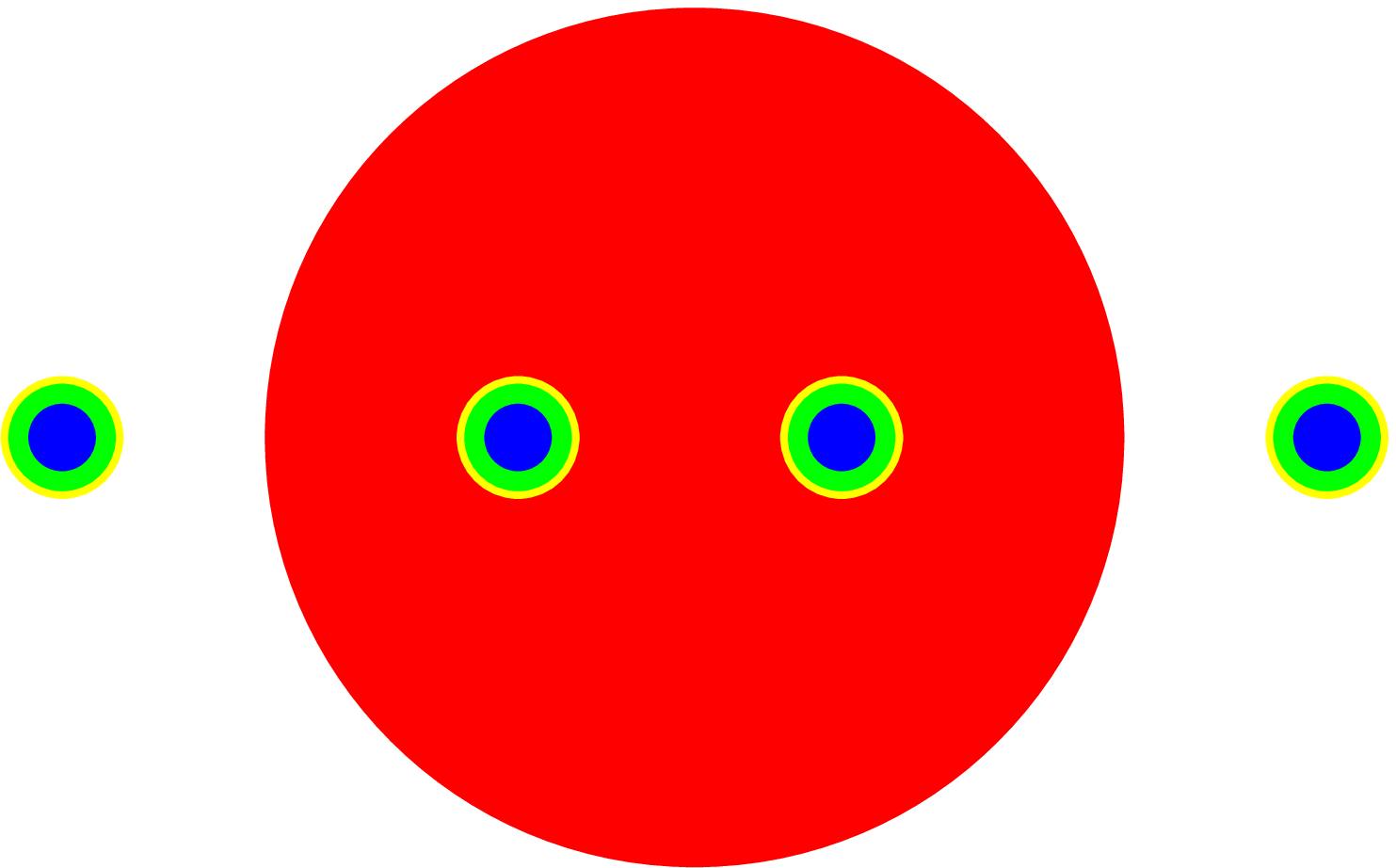}
}
\hspace{1cm}
\subfigure[$L_a=3$] 
{
    \label{fig:anti:b}
    \includegraphics[width=3cm]{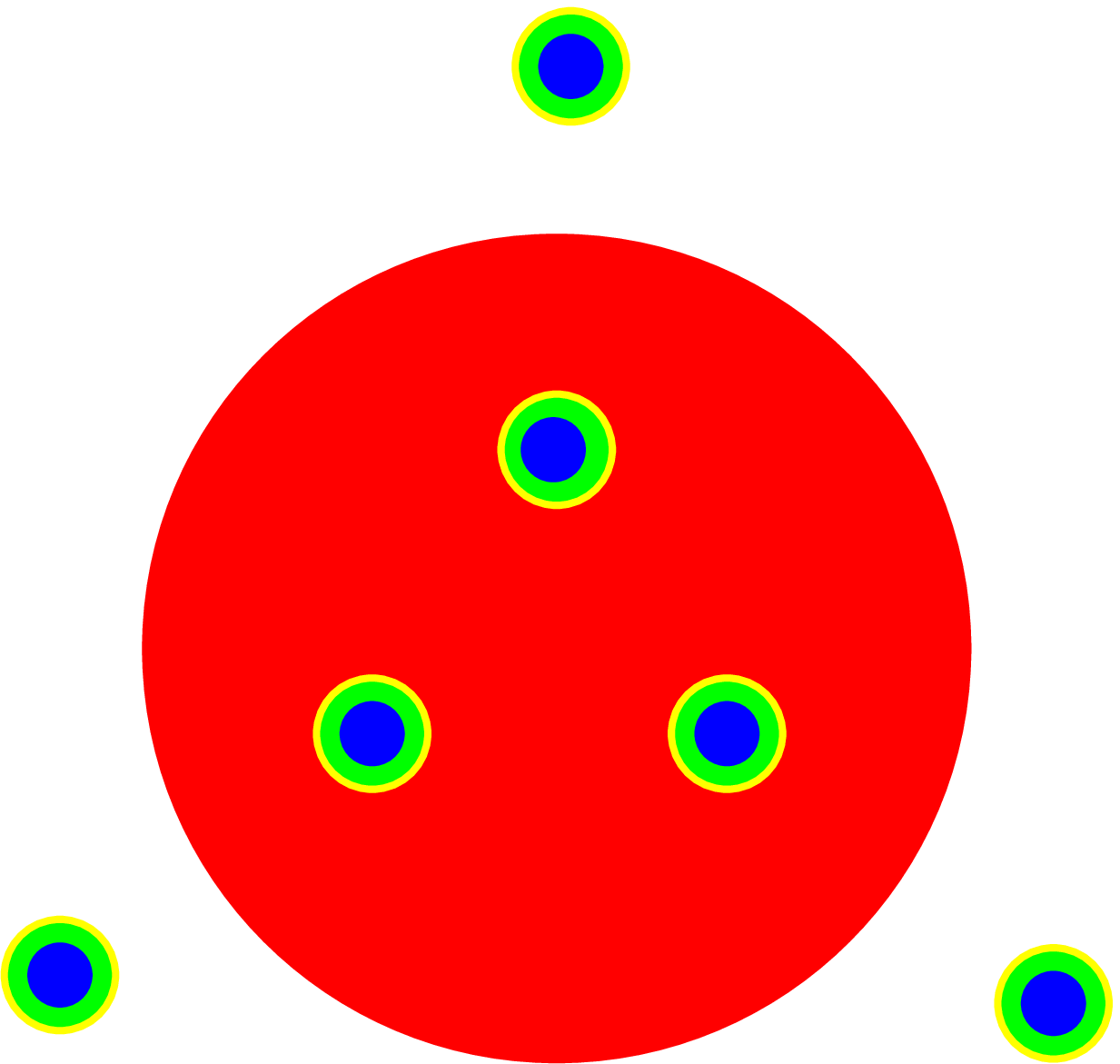}
}
\caption{Equilibrium configurations of $L_a=2$ and $L_a=3$ states.}
\label{fig:anti} %
\end{figure}

Next, we study SQF states in the presence of AV states. First, we optimize
Eq.(\ref{change}) with respect to positions of vortices and antivortices. 
In our numerical calculation, we start from, $R/\Lambda= 0.4$ and and 
try larger values of 
$R/\Lambda$. We limit ourselves to the $L_a=2$ and $L_a=3$ states.  
Equilibrium configurations of  
$L_a=2$ and $L_a=3$ states are depicted in Fig. \ref{fig:anti}. These 
configurations are independent of the dot's size, wheras equilibrium 
positions are dependent on $m \Phi_0/\varepsilon_0$. The further increase 
of this parameter causes the further push of vortices towards the dot's 
center, and antivortices away from the dot's boundaries.    As seen 
in Fig. \ref{fig:anti}, antivortices are aligned 
with vortices and appear on a ring outside the dot.  Next, we compare the 
effective energies of these states with SQF states without antivortices 
around them.  The comparison of 
effective energies with and without AV states shows that AV states are not 
energetically favorable. However, as discussed  in the previous section, 
for 
dots whose size is on the order of a few $\xi$, there might be stable 
AV states. However, the London theory does fail in that case. 

\end{multicols}

\begin{multicols}{2}

\section{Conclusions}

We studied the vortex states that occur spontaneously  in the ground states of
 a SC thin film with a single FM dot 
grown upon it, in the London approximation. Our calculations showed that 
GV states are more energetically favorable when the dot's size is up to  
$0.3 \Lambda$ ( $4.5 \xi $ ) . Between $R = 0. 3 \Lambda$ ( $4.5 \xi $ ) 
and 
$R = 0.4 \Lambda$ ( $ 6 \xi $ ) , both 
GV and SQF  states appear. Beyond $0.4 \Lambda$ ( $6 \xi $ ), only SQF 
states are 
stable. 

Furthermore,  
 we determine the geometric 
patterns of up to seven vortices. Our   results show that 
vortices 
form regular geometric patterns,  such as a regular triangle, square, 
pentagon and hexagon. However, in the case of seven vortices, one vortex 
occurs at the 
dots center, while other six sit at the corners of a hexagon. 
The phase diagram of SQF vortex states obeys an equation $N/(R/\Lambda)$, 
which suggests that high-vorticity states prefer larger dots.  

We also studied the cases together with AV states. However, our 
calculations did not show any stable vortex states together with 
antivortices. This result might seem implausable since one would expect 
the presence of AV states  due to the zero flux condition over the entire 
film. However, the zero flux condition can also be satisfied by the 
presence 
of a  supercurrent outside the dot boundaries, which circulates in 
the opposite 
direction to the currents created by vortices under the dot.
Furthermore, the London approach does not enable us to analyze the vortex 
and 
antivortex states  when 
the dot's size is as large as a few  $\xi$, because  it fails 
for distances on the order of coherence length. This failure is 
significantly transparant in the presence of AV states. However, positive 
vorticity states can be studied through the London approach. 

In closing, we studied SQF vortex , GV and AV states due to a magnetic 
dot in the  
London 
approximation. Experimentally, these states can be checked by magnetic 
force microscopy techniques.      

\section{Acknowledgments}

The author acknowledges Valery L. Pokrovsky for his very fruitful 
discussions.  
The most of this work was done during my stay at the University of 
Minnesota 
and was 
partially supported by the U.S. Department of Energy, Office of Science, 
under Contract No. W-31-109-ENG-38.

\end{multicols}

\end{document}